\documentclass[journal=jacsat,manuscript=article, layout=twocolumn]{achemso} 
\usepackage{blindtext}
\usepackage{graphicx} 
\usepackage{amsmath,siunitx}

\title{Realization of a Pre-Sample Photonic-based Free-Electron Modulator in Ultrafast Transmission Electron Microscopes}

\author{Beatrice Matilde Ferrari}
\affiliation[unimib]{LUMiNaD, Department of Materials Science, University of Milano-Bicocca, Milano, 20126, Italy}
\alsoaffiliation{LUMES, École Polytechnique Fédérale de Lausanne, Lausanne, 1015, Switzerland}

\author{Cameron James Richard Duncan}
\affiliation{LUMiNaD, Department of Materials Science, University of Milano-Bicocca, Milano, 20126, Italy}

\author{Michael Yannai}
\affiliation{Department of Electrical and Computer Engineering, Technion, Haifa, 32000, Israel}

\author{Raphael Dahan}
\affiliation{Department of Electrical and Computer Engineering, Technion, Haifa, 32000, Israel}

\author{Paolo Rosi}
\affiliation{CNR-Nano, Istituto di Nanoscienze Consiglio Nazionale delle Ricerche, Modena, 41125, Italy}

\author{Irene Ostroman}
\affiliation{LUMiNaD, Department of Materials Science, University of Milano-Bicocca, Milano, 20126, Italy}

\author{Maria Giulia Bravi}
\affiliation{LUMiNaD, Department of Materials Science, University of Milano-Bicocca, Milano, 20126, Italy}

\author{Arthur Niedermayr}
\affiliation{Department of Electrical and Computer Engineering, Technion, Haifa, 32000, Israel}

\author{Tom Lenkiewicz Abudi}
\affiliation{Department of Electrical and Computer Engineering, Technion, Haifa, 32000, Israel}

\author{Yuval Adiv}
\affiliation{Department of Electrical and Computer Engineering, Technion, Haifa, 32000, Israel}

\author{Tal Fishman}
\affiliation{Department of Electrical and Computer Engineering, Technion, Haifa, 32000, Israel}

\author{Sang Tae Park}
\affiliation{IDES-JEOL, Akishima, Tokyo, 196-8558, Japan}

\author{Dan Masiel}
\affiliation{IDES-JEOL, Akishima, Tokyo, 196-8558, Japan}

\author{Thomas Lagrange}
\affiliation{LUMES, École Polytechnique Fédérale de Lausanne, Lausanne, 1015, Switzerland}

\author{Fabrizio Carbone}
\affiliation{LUMES, École Polytechnique Fédérale de Lausanne, Lausanne, 1015, Switzerland}

\author{Vincenzo Grillo}
\affiliation{CNR-Nano, Istituto di Nanoscienze Consiglio Nazionale delle Ricerche, Modena, 41125, Italy}

\author{F. Javier García de Abajo}
\affiliation{ICFO-Institut de Ciences Fotoniques, The Barcelona Institute of Science and Technology, Castelldefels (Barcelona), 08860, Spain}
\alsoaffiliation{ICREA-Institució Catalana de Recerca i Estudis Avançats, Barcelona, 08010, Spain
}

\author{Ido Kaminer}
\affiliation{Department of Electrical and Computer Engineering, Technion, Haifa, 32000, Israel}

\author{Giovanni Maria Vanacore}
\affiliation{LUMiNaD, Department of Materials Science, University of Milano-Bicocca, Milano, 20126, Italy}
\email{giovanni.vanacore@unimib.it}

\begin{document}

\begin{abstract}
Spatial and temporal light modulation is a well-established technology that enables dynamic shaping of the phase and amplitude of optical fields, significantly enhancing the resolution and sensitivity of imaging methods.
Translating this capability to electron beams is highly desirable within the framework of a transmission electron microscope (TEM) to benefit from the nanometer spatial resolution of these instruments. 
In this work, we report on the experimental realization of a photonic-based free-electron modulator integrated into the column of two ultrafast TEMs for pre-sample electron-beam shaping.
Electron-photon interaction is employed to coherently modulate both the transverse and longitudinal components of the electron wave function, while leveraging dynamically controlled optical fields and tailored design of electron-laser-sample interaction geometry.
Using energy- and momentum-resolved electron detection, we successfully reconstruct the shaped electron wave function at the TEM sample plane.
These results demonstrate the ability to manipulate the electron wave function before probing the sample,  paving the way for the future development of innovative imaging methods in ultrafast electron microscopy.
\end{abstract}

\maketitle

\section{Introduction}

In recent years, electron-photon interaction (EPI) in Ultrafast Transmission Electron Microscopes (UTEMs) has been extensively adopted as a powerful tool for studying ultrafast dynamics of laser-triggered nanoscale phenomena \cite{vanacoreFourdimensionalElectronMicroscopy2016, barwickPhotonicsPlasmonics4D2015, zewailFourDimensionalElectronMicroscopy2010, arbouetChapterOneUltrafast2018, buttnerObservationFluctuationmediatedPicosecond2021, domroseNanoscaleOperandoImaging2024, trucLightInducedMetastableHidden2023, ungeheuerCoherentAcousticPhonons2024, tauchertPolarizedPhononsCarry2022, kazenwadelCoolingTimesFemtosecond2023, aguilarSelectiveProbingLongitudinal2023, wooEngineering2DMaterial2024}. 
Lately, there has been a growing interest in transferring well-established laser modulation techniques to the electron beam (e-beam) of UTEMs. In these experiments, pre-shaped laser fields transfer their modulation to the e-beam through EPI.\cite{vanacoreSpatiotemporalShapingFreeelectron2020, madanQuantumFutureMicroscopy2020} 
Several approaches, such as optical parametric amplification (OPA) \cite{bucherCoherentlyAmplifiedUltrafast2024},  two-wave mixing \cite{kozakInelasticPonderomotiveScattering2018}, as well as dielectric laser acceleration (DLA) \cite{adivQuantumNatureDielectric2021}, have been employed for longitudinal (energy-time) modulation of the electron wave packet, whereas spatial light modulators (SLMs)\cite{madanUltrafastTransverseModulation2022, chiritamihailaTransverseElectronBeamShaping2022}, nanoconfined near fields\cite{vanacoreUltrafastGenerationControl2019, tsessesTunablePhotoninducedSpatial2023}, and Fabry-P\'erot optical cavities \cite{schwartzLaserPhasePlate2019} have been used for transverse (momentum-space) shaping of the e-beam.

Two different approaches have been explored for e-beam modulation via light fields: elastic and inelastic interactions.
Elastic EPI, mediated by the ponderomotive force,  \cite{kozakPonderomotiveGenerationDetection2018, garciadeabajoOpticalModulationElectron2021, chiritamihailaTransverseElectronBeamShaping2022} offers the advantage of not requiring a physical interface to mediate the interaction. 
However, this method is constrained by the need for high laser intensities and is limited to phase-only modulation (unless two-color illumination is used to produce stimulated Compton scattering).
In contrast, inelastic EPI has been successfully used to modulate both the phase and amplitude of e-beams. A first step in that direction has been taken by exploiting the Photon-Induced Near-field Electron Microscopy (PINEM) approach \cite{barwickPhotoninducedFieldElectron2009, garciadeabajoMultiphotonAbsorptionEmission2010, parkPhotoninducedFieldElectron2010} for e-beam shaping \cite{vanacoreUltrafastGenerationControl2019, tsessesTunablePhotoninducedSpatial2023}. 
In PINEM, the interaction is mediated by the near field generated in the vicinity of a nanostructure (such as a surface plasmon polariton), allowing direct mapping of the structure’s complex optical field onto the transverse distribution of the electron wave function. While this one-to-one correspondence is valuable for studying the near-field evolution, it limits the flexibility in e-beam shaping: predicting the final electron distribution requires numerical simulations to account for the near-field effects of the structure, limiting the range of patterns that can be imprinted on the e-beam.

To overcome these limitations, stimulated Inverse Transition Radiation (ITR) \cite{steinhauerInverseTransitionRadiation1997, weingartshoferDirectObservationMultiphoton1977, plettnerVisibleLaserAccelerationRelativistic2005} has emerged as a versatile alternative form of EPI \cite{ vanacoreAttosecondCoherentControl2018, morimotoDiffractionMicroscopyAttosecond2018, feistHighpurityFreeelectronMomentum2020, madanUltrafastTransverseModulation2022, wangEnergymomentumTransferFreeelectronphoton2024, fangStructuredElectronsChiral2024}. Transition radiation (TR) is the electromagnetic wave emitted by an electron passing through an interface between two media with different refractive indices. 
This radiation is emitted to preserve the continuity of electromagnetic fields at the interface \cite{glauberQuantumOpticsDielectric1991}.
In the inverse process (i.e., ITR), a stimulating electromagnetic field is already present, and the electron exchanges energy and momentum with such a field in a quantized manner.
As a result, ITR allows for direct transfer of the laser phase onto the electron wave function without the near-field contribution present in PINEM, making the final e-beam profile effectively a Fourier transform of the SLM pattern used to shape the laser \cite{konecnaSinglePixelImagingSpace2023}. 


Recently, Madan \textit{et al.} \cite{madanUltrafastTransverseModulation2022} demonstrated electron modulation mediated via inelastic EPI based on ITR using light pre-shaped by an SLM. In that work, the interaction was still occurring at the sample plane, where the film used to mediate ITR was placed. Nevertheless, the authors foresaw a future technological implementation of a pre-sample Photonic-based free-ELectron Modulator (PELM). This would be done by adding a new EPI interaction point in a pre-sample stage along the TEM column. Different types of modulation can be then achieved by exploiting the multiplicity of phase patterns that can be imprinted on the SLM, as well as by replacing the SLM itself with other laser shaping technologies, such as optical cavities, DLAs, OPAs, leveraging both spatially and temporally modulated light fields for controlling the e-beam in its multidimensional phase space before reaching the sample.

Realizing such potentials will allow us to use pre-shaped e-beams to selectively probe nanomaterial dynamics with enhanced spatiotemporal resolution and sensitivity to specific properties and degrees of freedom. For instance, shaped electrons could be adopted to selectively probe low-frequency excitations in materials \cite{fangStructuredElectronsChiral2024, mattesFemtosecondElectronBeam2024, gaidaAttosecondElectronMicroscopy2024}, to enable low-dose imaging of sensitive scatterers \cite{konecnaSinglePixelImagingSpace2023}, as well as to enhance image resolution \cite{garciadeabajoOpticalModulationElectron2021} and increase contrast \cite{bucherCoherentlyAmplifiedUltrafast2024} when interrogating materials showing very subtle changes, thus greatly expanding the capabilities of UTEMs. 

In this work, we present the experimental implementation of a PELM integrated into two UTEM systems: one at the University of Milano-Bicocca (UniMiB) and the other at the Israel Institute of Technology (Technion).  Both setups were modified to position the PELM before the sample, although at different locations along the TEM column. At UniMiB, the PELM utilizes a SLM for transverse e-beam modulation, while at Technion, a non-collinear OPA (NOPA) is used for longitudinal modulation. By monitoring the electron wave function in its multidimensional phase space (energy, time, space, and momentum), we demonstrate both transverse and longitudinal pre-sample modulation of the e-beam, driven by an externally-controlled light field.

\section{Results and discussion}

\subsection{Electron-photon interaction and coherence}

As described above, our PELMs exploit ITR to facilitate EPI. In this process, electrons inelastically interact with a strong coherent light source, absorbing and emitting an integer number of photons \cite{giulioProbingQuantumOptical2019}.
This quantized interaction imprints distinct peaks in the electron distribution, corresponding to multiples of the photon’s energy and momentum \cite{weingartshoferDirectObservationMultiphoton1977, plettnerVisibleLaserAccelerationRelativistic2005, vanacoreAttosecondCoherentControl2018, feistHighpurityFreeelectronMomentum2020, wangEnergymomentumTransferFreeelectronphoton2024, yangUnifyingFrequencyMetrology2024}.

To resolve these interaction peaks, and hence the quantized nature of EPI, the e-beam energy (or momentum) spread must be narrower than the photon’s energy (or momentum). Such narrow spreads are achieved when the e-beam longitudinal (or transverse) coherence length exceeds the wavelength of the interacting photons \cite{baumPhysicsUltrashortSingleelectron2013}.
This is because the longitudinal and transverse coherence lengths directly determine the e-beam spread in energy and momentum, respectively, when averaging over the electron ensemble of wave functions \cite{fowlesIntroductionModernOptics1989}. It is important to notice that, if the coherence length exceeds the photon wavelength but remains on the same order of magnitude, it is still possible to detect EPI; however, the interaction peaks will be smeared out and the quantum nature of EPI hidden. 

In the energy domain, EPI is detected if the temporal coherence of the e-beam is longer than or on the order of an optical cycle. This condition is easily achieved thanks to the small energy spread that results from the photoemission process at the cathode \cite{janzenPulsedElectronGun2007}. As a consequence, the electron pulse has a temporal coherence $\xi_t \approx 5-\SI{10}{fs}$, which is greater than the optical cycle (in our case, $\tau = \lambda/c \approx \SI{3.4}{fs}$) \cite{baumPhysicsUltrashortSingleelectron2013}.

In contrast, detecting EPI in the momentum domain is more challenging. In this case, the electron transverse coherence must be greater than or on the order of the laser wavelength. 
Typical values of transverse coherence for thermionic electron sources, such as those used in our laboratories, are on the order of tens of nanometers \cite{kirchnerCoherenceFemtosecondSingle2013, gahlmannUltrashortElectronPulses2008}, which is not enough for experiments with visible and infrared (IR) light. Therefore, particular microscope settings need to be used to increase the beam transverse coherence, as discussed below.

For e-beam shaping, it is crucial not only to achieve sufficient e-beam coherence but also to achieve coherent EPI, since preserving the phase information is critical for advanced imaging techniques. 
Coherent EPI occurs when the e-beam interacts with a homogeneous portion of the laser's electric field.
For longitudinally coherent EPI, the laser pulse is stretched in time to exceed the duration of the electron pulse \cite{vanacoreAttosecondCoherentControl2018, feistHighpurityFreeelectronMomentum2020}.
For transversely coherent EPI, the e-beam spot size is made smaller than the laser spot size \cite{feistHighpurityFreeelectronMomentum2020}.

\subsection{Technical Implementation of the PELM device}

To realize EPI, we implement a pump-probe scheme. An IR laser pulse, produced by an Yb-based amplified femtosecond laser, is split in two branches. One pulse is up-converted to ultraviolet (UV) light by a 4th harmonic generation stage and is directed to the TEM cathode, inducing photoemission and creating the probing electron pulse. 
The other pulse is synchronized with the electron pulse and pre-shaped via the SLM. 
The two pulses interact via ITR on a light-reflective, electron-transparent metallic film at the PELM plane, where the structured laser field imprints its modulation onto the electron wave function.
To maximize EPI, the shaped laser pulse is p-polarized with respect to the PELM film \cite{madanUltrafastTransverseModulation2022}.

\begin{figure*}[!ht]

    \includegraphics[width=12cm]{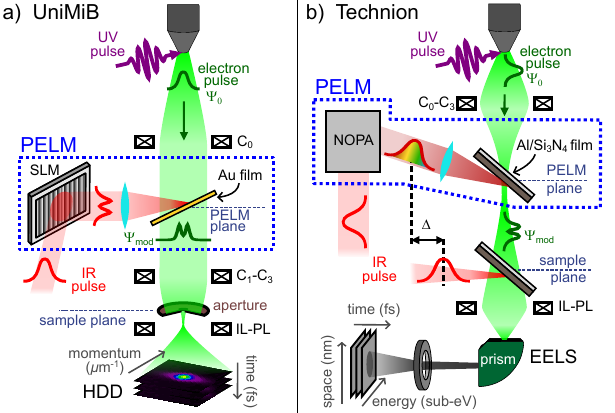}
    \caption{\textbf{Schematics of experimental setups incorporating the Photonic Electron Modulator (PELM) in two ultrafast TEM configurations} at University of Milano-Bicocca (UniMiB) (a) and the Israel Institute of Technology (b). 
    \textbf{a) Transverse electron-beam shaping at UniMiB}. A Spatial Light Modulator (SLM) is used to shape the transverse laser profile, which is then focused on a thin gold film positioned in a pre-condenser-lens (pre-CL) stage inside the TEM. Femtosecond electron pulses generated by UV laser pulses interact with the modulated IR laser pulses at the PELM-film surface via stimulated inverse transition radiation (ITR).
    A 20-$\mu$m aperture selectively samples a small portion of the modulated e-beam. High Dispersion Diffraction (HDD) patterns of the selected electrons are recorded. 
    By scanning the aperture or laser, the entire transverse electron profile is reconstructed, mimicking 4D-STEM methodologies.
    \textbf{b) Longitudinal electron-beam shaping at Technion.} A Non-collinear Optical Parametric Amplifier (NOPA) is used to tune the optical cycle of the upper laser pulse, which is then directed onto a thin aluminum film deposited on a Si\textsubscript{3}N\textsubscript{4} membrane located at the Hard X-ray Aperture (HXA) position, between the condenser and objective lenses. The electron pulses are modulated via ITR at the PELM stage.
    The longitudinal electron modulation is then probed via a double EPI scheme with additional ITR interaction point at the sample plane. 
	}
\label{fig1}
\end{figure*}

The technical implementation of a PELM device requires integrating two pre-sample access ports to the TEM column reaching the same inner position: an optical port for the modulating laser beam and a micromanipulator to adjust the stage holding the PELM film.
This setup must satisfy two main design constraints: (i) sufficient lateral and vertical space within the column to accommodate the PELM film, and (ii) adequate transverse coherence of the e-beam to enable EPI detection.
Here, we devise two configurations for integrating the PELM within UTEM setups, each with unique advantages and trade-offs.

Figure~\ref{fig1}a shows the UTEM setup at UniMiB, with more details provided in the Supporting Information (SI). This setup is based on a modified JEOL JEM-2100 TEM operating at \SI{200}{keV} and equipped with a direct electron detector. The microscope column has been modified to accommodate two additional column sections before the C$_1$-C$_3$ condenser lenses (CLs). The upper section houses a supplementary condenser lens, labeled C$_0$, while the lower section hosts: a single optical port for both UV and IR beams, an aluminum mirror to guide the UV beam toward the cathode, and the micromanipulator holding the PELM film (see Fig.~S1).

Because of its pre-CL position, the PELM at UniMiB offers the key advantage of controlling the light-shaped e-beam via the CLs before reaching the sample. 
This pre-CL shaping is essential for implementing advanced imaging techniques that require, for instance, demagnification of the modulated e-beam at the sample \cite{konecnaSinglePixelImagingSpace2023}.
However, achieving optimal transverse coherence at this pre-CL stage is challenging; under typical imaging conditions, this is generally accomplished by demagnification with the CLs and transverse beam selection with the condenser aperture.
Here, to reach the coherence required for resolving EPI at the pre-CL PELM plane, we substantially reduce the excitation voltage of the C$_0$ lens (see SI) and we use the PELM stage itself as an aperture. Consequently, the e-beam diameter significantly exceeds that of the laser, resulting in substantial loss of useful electron flux, as many electrons fall outside the laser interaction region.

Figure~\ref{fig1}b illustrates the UTEM setup at Technion \cite{bucherCoherentlyAmplifiedUltrafast2024} (a real picture is also shown in Fig.~S6), based on a modified JEOL JEM-2100Plus TEM operating at \SI{200}{keV} and equipped with a Gatan Image Filter and a K2 direct electron detector. 
This configuration includes a single additional column section with: the C$_0$ lens, the aluminum mirror for the UV beam, and the UV optical port.
In this setup, the PELM film is accommodated by modifying the Hard-X-Ray aperture (HXA), located between the condenser and objective lenses. 
The IR-beam access to the PELM stage is provided through an optical port on the opposite side of the column, with an entrance angle of $20^{\circ}$ relative to the horizontal plane.

At Technion, an optical pump line to the sample is implemented by further splitting the IR beam in two paths (see panel b of Fig.~S6 in SI). One path is directed to the sample plane via a zero-angle port, while the other enters the microscope at the PELM port, after passing through a fine-delay line.

With respect to the pre-CL configuration implemented at UniMiB, the advantages of the post-CL configuration at Technion are: (i) a lower complexity in terms of technical design and practical realization, because the standard configuration of a TEM column is already designed to host a HXA; and (ii) a higher e-beam transverse coherence, which is ensured by the CL system.
The latter aspect has been qualitatively investigated via electron trajectory calculations using the STEM-CELL software \cite{grilloSTEM_CELLSoftwareTool2013, grilloSTEM_CELLSoftwareTool2013a} (see related section in the SI). 

However, such post-CL PELM is limited by the tighter space available in this portion of the column, thus suffering from a lower flexibility in controlling the modulated e-beam before the sample and therefore limiting its versatility.

\subsection{Pre-sample electron beam shaping}

Having discussed the technical design of the PELM devices, we now proceed to demonstrate their ability to control the e-beam properties within their multidimensional phase space.

\subsubsection{Transverse modulation}
Transverse modulation of the e-beam requires modifying both the spatial and momentum coordinates of the single-electron wave function. This can be achieved using spatially structured optical fields. In our case, we employ an SLM to shape the light field, but alternative approaches, such as nanoconfined near fields or photonic cavities, could serve a similar purpose.

\begin{figure*}[!h]
    \includegraphics[]{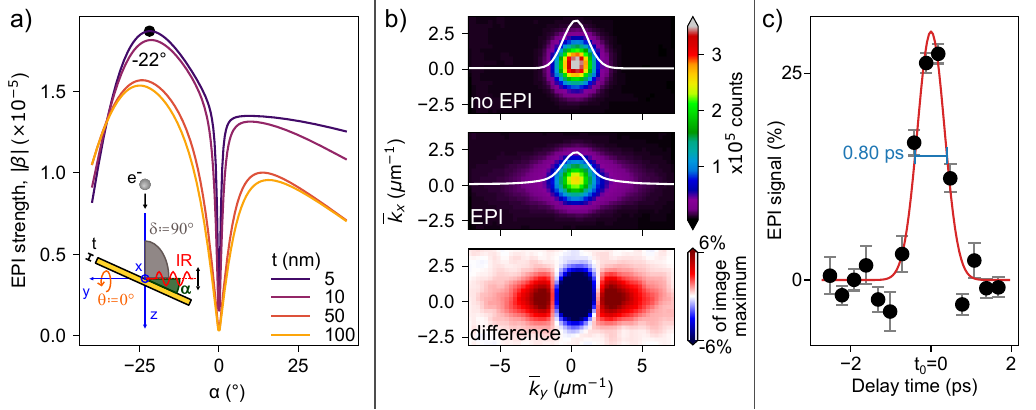}
    \caption{\textbf{Analysis of EPI at UniMiB}. \textbf{a)} Simulated interaction strength as a function of the gold-film tilt angle $\alpha$, accounting for non-perfect mirror conditions.  The maximum EPI strength occurs at $\alpha = -22^\circ$, which is used in all subsequent measurements. The inset illustrates the experimental geometry of EPI at the PELM film. \textbf{b)} High Dispersion Diffraction (HDD) patterns of the e-beam under different conditions: without laser illumination (top panel), after interaction with IR photons (central panel), and their difference (lower panel). The interaction broadens the electron wave function along the $k_y$ direction (laser propagation axis) and redistributes the intensity, reducing the signal near $k = 0$ and increasing it at higher $k_y$. \textbf{c)} Temporal evolution of the EPI signal (defined in the main text) as a function of the laser-electron pulse delay. Black dots are data and red-solid line is a Gaussian fit. 
    The extracted full width at half maximum (FWHM) is $800 \pm \SI{30}{fs}$. We define the temporal overlap $t_0$ between the two pulses as the center of the Gaussian.
	}
\label{fig2}
\end{figure*}

At UniMiB, we achieve light-induced transverse e-beam modulation by exploiting inelastic EPI via ITR, which is mediated by a 5-nm thick gold film at the PELM plane. This thickness is not sufficient to approximate the PELM film with a perfect mirror, given that the skin depth of gold is around $t=\lambda/(2\pi \sqrt{\varepsilon_2/2})\approx27\,$nm, where $\lambda = \SI{1030}{nm}$ is the wavelength of our laser. 
Using the approach outlined in the SI of Ref. \citenum{vanacoreAttosecondCoherentControl2018}, we simulated the strength of EPI --- called $\beta$ or $g$ in the literature --- as a function of the PELM film tilt angle $\alpha$ (as shown in the inset of Fig.~\ref{fig2}a), accounting for the non-ideal mirror conditions. 
Our results, displayed in Fig.~\ref{fig2}a, indicate that an optimal interaction strength occurs at a tilt angle of $\alpha = -22^\circ$, which was therefore used in all following experiments.

Here, we measure EPI by imaging the transverse momentum distribution of the e-beam in High Dispersion Diffraction (HDD) mode using a 100-m camera length on a direct electron detector (see SI for further information). 
The large camera length is crucial to resolve small momentum transfers.
Indeed, the interaction imparts discrete multiples of the light film-projected momentum $k_L=(2\pi/\lambda) \cos^2(\alpha)=\frac{2\pi}{\SI{1030}{nm}} \cos^2(22^\circ) \approx \SI{5.2}{\mu m^{-1}}$ (see also SI), which is equivalent to an angular deflection of $\theta \approx (k_\perp/k_\parallel)=k_L/(2 \pi /\lambda_e) \approx \SI{2.1}{\mu rad}$, where $\lambda_e \approx \SI{2.5}{pm}$ is the relativistic de Broglie wavelength for 200-keV electrons.

Figure~\ref{fig2}b presents HDD momentum patterns of the e-beam under different conditions. In the absence of EPI (no light), the e-beam has a Gaussian momentum distribution (top panel). In the central panel, following interaction with IR photons, the electron wave function broadens along the $k_y$-axis, corresponding to the laser propagation direction \cite{vanacoreAttosecondCoherentControl2018} (see SI).

The lower panel in Fig.~\ref{fig2}b displays the difference between the upper two panels, enhancing the contrast to clearly highlight the effect of EPI. 
The interaction reduces the electron signal around $k = 0$ while increasing it at higher $k_y$ values.
The absence of resolved stripes at multiples of $k_L$ --- as the ones shown in Fig.~\ref{technionEnergyMap}b --- suggests that the transverse coherence length of the e-beam is on the order of the laser wavelength, limiting our ability to spatially resolve the quantum nature of the interaction.

We performed the experiment as a function of the delay time between laser and electron pulse to achieve a precise temporal overlap between the two ($t_0$: delay time = \SI{0}{ps}). The results are shown in Fig.~\ref{fig2}c. 
The EPI signal is represented by the depletion of the direct e-beam and we quantify it as $1 - A_{\mathrm{Voigt}}$, where $A_{\mathrm{Voigt}}$ is the amplitude of a Voigt function fitted to the normalized $k_y$-integrated pattern.
The temporal evolution of the EPI signal reveals a full width at half maximum (FWHM) of $800 \pm \SI{30}{fs}$, consistent with previous studies \cite{barwickPhotonicsPlasmonics4D2015}.

\begin{figure*}[!htb]
	\includegraphics[]{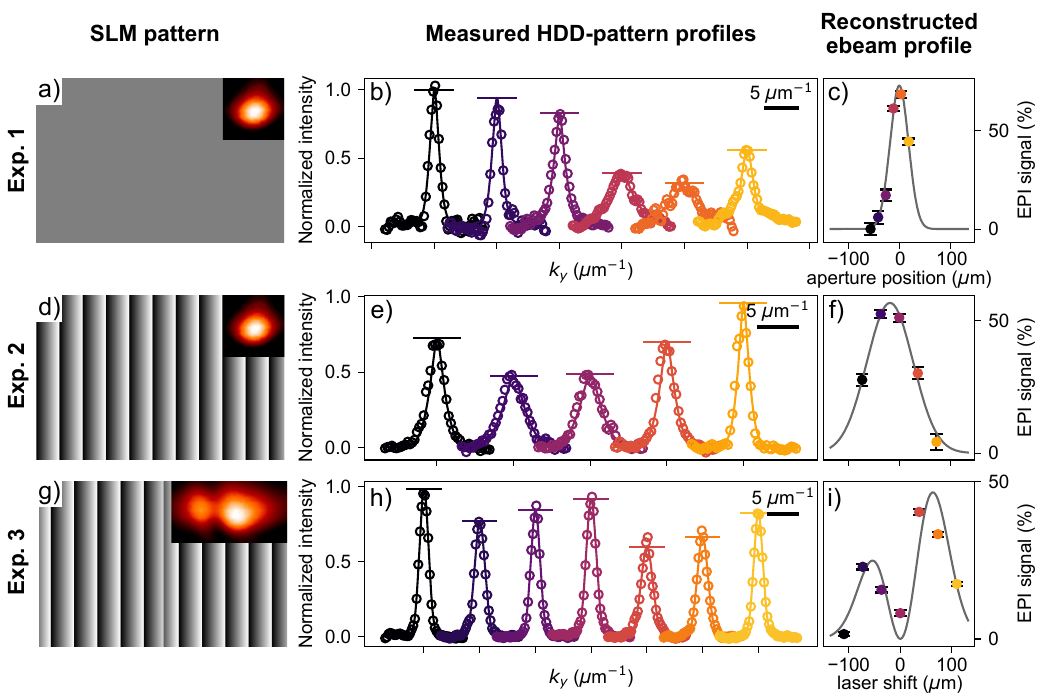}
	\caption{\textbf{Reconstruction of electron-beam transverse profiles at UniMiB.} \textbf{a,d,g)} patterns displayed by the SLM. The color scheme consist of a gray level of the pixels going from 0 to 255 to represent a laser phase shift from 0 to mod $2\pi$. 
    The insets show the modulated laser transverse profile recorded with a CCD camera in the conjugate plane (see SI for further details on the experimental setup). \textbf{b,e,h)} $k_x$-integrated profiles (circles) of the HDD patterns acquired at maximum EPI. 
    The colors represent different aperture positions (b) or laser shifts (e,h) coordinated with panels c,f,i). 
    The curves are laterally shifted for clarity. Solid curves are Voigt-function fits while horizontal lines represent the fitted amplitudes $A_{\mathrm{Voigt}}$.
    \textbf{c,f,i)} Electron-photon interaction (EPI) signal, $1-A_{\mathrm{Voigt}}$, as a function of the aperture position (c) or the laser shift (f,i). Dot colors in these panels match the corresponding profiles in panels b,e,h. The gray line is a gaussian fit (c,f) or a representative profile of the Hermite-Gaussian 01 mode.}
\label{unimibSpatial}
\end{figure*}

Having demonstrated the ability to detect the effect of EPI on the electron wave function in momentum space at UniMiB, we performed 4D-STEM experiments to demonstrate transverse spatial modulation of the e-beam.
In 4D-STEM, a focused e-beam is scanned across the sample while simultaneously capturing at each scan position a convergent-beam electron diffraction pattern, providing phase contrast analysis of the material under investigation \cite{ophusFourDimensionalScanningTransmission2019, koutensky_ultrafast_2025}.
Here, a focused light beam is scanned across the e-beam on the PELM film while simultaneously capturing at each scan position an HDD electron diffraction pattern (see also Fig.~\ref{fig1}), providing transverse phase contrast analysis of the shaped e-beam.

Figure~\ref{unimibSpatial} illustrates the experiments conducted with this 4D-STEM approach, with the electron and laser pulses in temporal overlap. 
A 20-$\mu$m aperture (as the ones commonly used in a TEM column) is placed in the sample holder, downstream of the PELM stage (see Fig.~\ref{fig1}). 
This aperture, smaller than the laser spot size, selects the electrons that interacted with a homogeneous portion of the electromagnetic field, condition needed for coherent interaction \cite{feistHighpurityFreeelectronMomentum2020}.
Electrons passing through the most intense regions of the laser beam interact with more photons, resulting in a stronger EPI signal, and vice versa.

The first row of Fig.~\ref{unimibSpatial} depicts the initial experiment, in which the SLM serves as a simple mirror (pattern shown in panel a); the laser transverse profile is Gaussian, as shown in the inset.  
The aperture is scanned across this laser profile, and for each position, an HDD electron pattern is taken (as in Fig.~\ref{fig2}b) before and at $t_0$ (see Fig.~\ref{fig2}c).
The pre-$t_0$ images serve as reference to normalize the $k_x$-integrated profiles measured at maximum interaction. 
The resulting profiles, shown as circles in panel b, are fitted with a Voigt function (solid curves in panel b) and the fitted amplitudes are displayed as horizontal lines. 
Panel c shows the EPI signal, $1-A_{\mathrm{Voigt}}$, as a function of the aperture position, effectively reconstructing the e-beam profile at the aperture (sample) plane. We estimate the Gaussian-modulation diameter of the e-beam (at $1/e^2$ intensity) to be approximately $72 \pm \SI{5}{\mu m}$.

The second row of Fig.~\ref{unimibSpatial} presents a similar experiment where the aperture remains fixed while the laser beam is scanned across the PELM film using the SLM.
In this case, a blazed grating (shown in panel d) is employed to shift the laser beam in a controlled manner, based on the diffraction condition for the first order:  $\Delta \delta = \lambda\Delta n/H$. Here, $\Delta \delta$ is the laser tilt change, $H$ is the SLM horizontal dimension and $\Delta n$ is the difference in the number of grating periods, which is varied to scan the laser on the PELM stage (see also SI). 
From these calculations, we can estimate the Gaussian modulation diameter at the PELM plane to have a $1/e^2$-diameter of $180 \pm \SI{20}{\mu m}$.
By comparing panel f with panel c, we deduce that the e-beam has been de-magnified by a factor $2.5$ between the PELM and the sample plane, consistent with the almost-parallel configuration of the e-beam post-PELM (see SI) and verifying that experiment 1 and experiment 2 give the same result.

The third row of Fig.~\ref{unimibSpatial} depicts a further variation of the experiment, using a different SLM pattern. Here, a horizontal phase shift is superimposed on the blazed grating pattern (panel g),  producing a two-lobed Hermite-Gaussian profile HG$_{01}$ in the far field, as shown in the inset of panel g. The integrated HDD patterns in panel h allow reconstruction of the e-beam profile (panel i), which mirrors the two-lobed shape imparted by the light beam.

Our results clearly demonstrate the ability to manipulate the e-beam prior to sample interaction and to observe the resulting modulations at the sample plane. These findings demonstrate that such a PELM device is now ready for performing ultrafast measurements with a shaped e-beam on real samples. Furthermore, our system demonstrates versatility, not only in e-beam shaping but also in e-beam demagnification, as we have shown by reducing the e-beam pattern by a factor of 2.5, with further scalability possible.

\subsubsection{Longitudinal modulation}

Influencing the longitudinal phase of the e-beam requires temporal modification of its single-particle wave function. Because time modification beyond free evolution directly implies modification of the energy spectrum, we can access the electron longitudinal phase shift via direct imaging in the energy domain. From a technological point of view, appropriate time-dependent optical fields, as obtained via OPAs (used in this work), two-wave mixing, or DLAs, are needed in order to achieve the desired modulation.

At Technion, we achieve light-induced longitudinal e-beam modulation by exploiting inelastic EPI via ITR, which is mediated by a 30-nm-thick Aluminum film deposited on a 40-nm Si$_3$N$_4$ membrane. In this case, the metallic film closely behaves as a perfect mirror.

\begin{figure}[!ht]
    \includegraphics[]{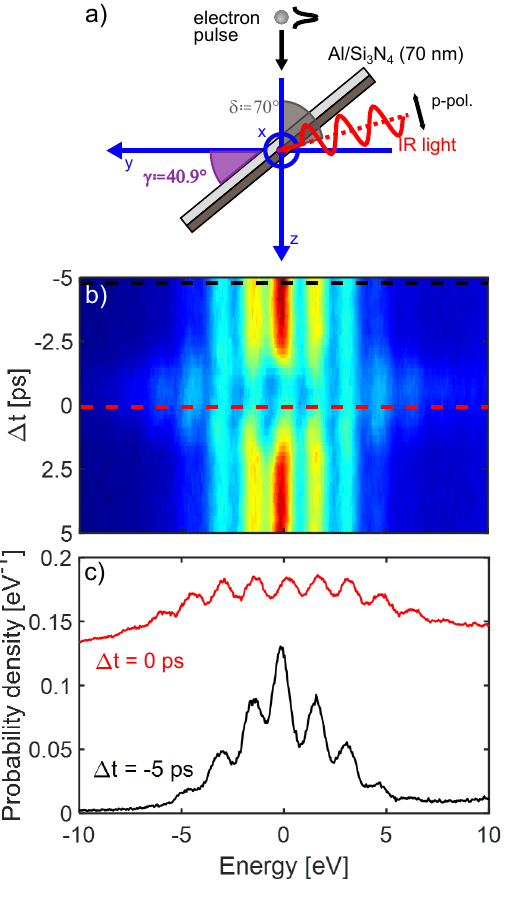}
    \caption{\textbf{Analysis of double EPI at Technion.} \textbf{a)} Experimental geometry of EPI at the PELM film. \textbf{b)} Electron energy spectra versus PELM time-delay taken while the sample laser is at optimal temporal overlap, thus demonstrating {\it double} EPI. \textbf{c)} Cross-sections taken from the data in panel a along the color-coordinated dashed lines, showing the EPI spectra at $\Delta t = \SI{-5}{ps}$ (only sample EPI), and $\Delta t = \SI{0}{ps}$ (both sample and PELM EPI).
    }
\label{technionEnergyMap}
\end{figure}

For longitudinal modulation to be adopted in imaging, it is crucial that the light field imprints the same phase shift at every transverse position of the electron wave function. 
In order to assure exact phase matching at the PELM plane between electrons and light interacting at the PELM-film interface, the film is tilted by an angle $\gamma = 40.9^{\circ}$ with respect to the horizontal plane (shown in Fig.~\ref{technionEnergyMap}a).
This angle can be obtained as: $\tan(\gamma) = \sin\delta/(c/v- \cos\delta)$, where $\delta$ is the angle between the electron and light (fixed at $70^{\circ}$), $c$ is the light speed and $v$ is the electron speed ($v = 0.7c$ at 200 keV). 
Such expression is derived by imposing that the phase shift experienced by the electron pulse while interacting with the light field ($\Delta\Phi_{e} =(\omega\Delta z/v)\sin\gamma$, where $\Delta z$ is the interaction distance), is equal to the projected phase of the light pulse on the PELM-film surface ($\Delta\Phi_{L} = -\Delta z\frac{\omega}{c}\cos(\pi/2 - \delta + \gamma)$). 
This condition makes sure that a spatially-uniform longitudinal phase profile is imprinted on the electron wave function at the PELM plane.

Here, we measure EPI by monitoring the energy distribution of the e-beam via the acquisition of either energy-resolved spectra or energy-filtered images. 

\begin{figure*}[!ht]
	\includegraphics[]{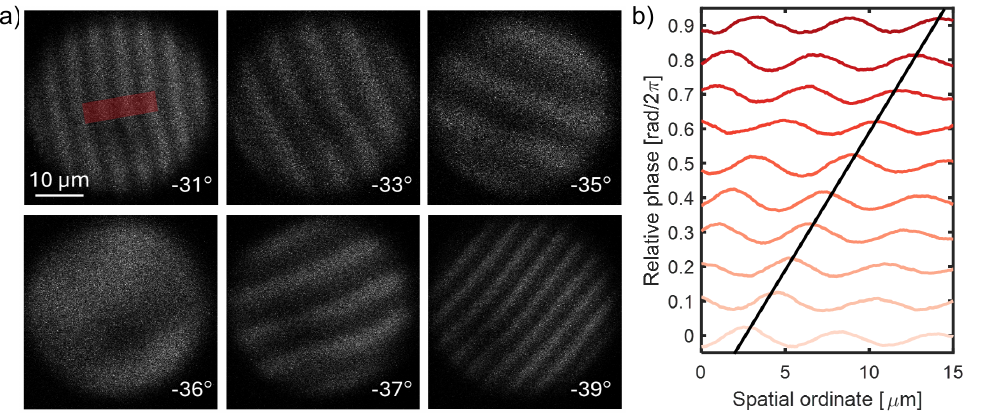}
	\caption{
    \textbf{Electron Interferometry and Phase Matching via Double EPI at Technion. a)} Electron interferograms obtained by recording energy-filtered TEM images for different sample tilt angles, and at optimal temporal overlap at both the PELM and the sample. Sample tilt angle is indicated for each panel. The bottom-left panel ($-36^\circ$) shows near-perfect phase matching (0 to $2\pi$ phase modulation across the ROI). \textbf{b)} Cross-sections of an interferogram, similar to those shown in a, taken while sweeping the relative PELM-sample time-delay (or phase) across a full optical period (\SI{2.73}{fs}). The black line traces the position of one peak for varying delays. The red rectangle in the top-left panel of a illustrates the ROI used to draw the cross-sections in b.
	}
\label{TechnionPhase}
\end{figure*}

In Figs.~\ref{technionEnergyMap} and \ref{TechnionPhase} we show the results of the simultaneous illumination of two Al/Si$_3$N$_4$ films, one placed at the PELM plane and tilted by $\gamma = 40.9^{\circ}$, and the other one placed at the sample plane with a variable tilting angle. Such configuration results in two subsequent EPIs both governed by ITR mechanism.

In Fig.~\ref{technionEnergyMap}b we observe the effect of such double interaction in the energy-time dimensions.
Electron energy loss spectra are measured as a function of the delay time, $\Delta t$, between the modulation pulse and the pump pulse, while keeping the electron pulse and the light pulse at the sample always at their optimal temporal overlap. 
At $\Delta t$ values very different from zero (black linecut in Fig.~\ref{technionEnergyMap}c), the spectrum shows multiple quantized energy exchanges on both sides of the zero-loss peak (ZLP) as a result of EPI only at the sample stage. However, while approaching the optimal temporal overlap also between electron and light pulses at the PELM stage (red linecut), a stronger modulation across the spectrum is observed especially in energy regions far away from the ZLP. 

A direct evidence of the coherent phase modulation experienced by the electron wave packet can be obtained via energy-filtering imaging of the spatial distribution of the e-beam following such double interaction.
In Fig.~\ref{TechnionPhase} we present the results of such imaging experiments, where energy-filtered images have been acquired as a function of the tilting angle of the sample film while maintaining optimal temporal overlap between electrons and light at both PELM and sample stages. Here, it is possible to observe the formation of a series of spatial fringes induced by the coherent superposition of the electron wave functions modulated by the two interaction points (PELM and sample). The periodicity of the fringes depends on the projection of the light wave vector on the tilted sample film and thus strongly changes when varying the sample tilting angle.

The short separation between the PELM position and the sample position is ideal for dispersive reshaping of the electron pulse to obtain attosecond longitudinal modulation of the beam in a Ramsey-like setup \cite{bucherCoherentlyAmplifiedUltrafast2024, gaidaAttosecondElectronMicroscopy2024, nabbenAttosecondElectronMicroscopy2023}.
In this respect it is crucial to have a precise correlation between the phase shift imprinted on the electron wave function and the relative phase difference between the two light pulses, the one driving the e-beam modulation and the other driving the sample excitation. 
In Fig.~\ref{TechnionPhase}a we show electron interferograms measured at a given sample tilting angle as a function of the delay time $\Delta t$ between the PELM laser and the pump (sample) laser within a single optical cycle. 
At each delay time, the measured shift of the electron fringes in the energy-filtered images is correlated with the optical phase difference (Fig.~\ref{TechnionPhase}b) as measured via a Mach-Zender interferometer. The latter is shown in details in Fig.~S6b of the SI: the laser beams directed at the PELM and sample planes are further separated and delayed with respect to each other before arriving on a CCD camera,  where optical interference is recorded.
From the measurements, we clearly observe a strong correlation between the optical phase and the electron phase, confirming that the PELM device is properly behaving.

\section{Conclusions}

In conclusion, we have successfully integrated a PELM device into the UTEMs at UniMiB and Technion, enabling on-demand control over the electron wave function in both longitudinal and transverse directions. 
This was achieved through the integration of an additional EPI stage within the TEM column, allowing for e-beam modulation before the sample plane. This novel configuration expands the possibilities for using shaped e-beams in ultrafast electron microscopy experiments. 

Given the capabilities of tailored e-beam shaping, PELM-equipped UTEMs open up new avenues for image-resolution enhancements, selective probing, and low-dose imaging. In particular, potential applications include single-pixel electron imaging, where a high number of patterns have to be imprinted on the electron wave function to obtain high-resolution images with minimal electron dose \cite{konecnaSinglePixelImagingSpace2023}. 
This capability could revolutionize imaging techniques, making the PELM system an essential tool for the next generation of ultrafast electron microscopy studies.

\section{Associated content}
The Supporting Information provides detailed technical descriptions and geometry of the UTEM setup at the University of Milano-Bicocca, including the phase-modulating Spatial Light Modulator.
Calibration procedures for lens configurations (e.g., C$_0$ and C$_3$ lenses) and electron beam coherence optimization are explained, alongside calculations for camera length and momentum transfer during electron-photon interactions. 
The document also describes the role of apertures in controlling beam coherence and intensity, compares pre- and post-condenser lens PELM configurations, and discusses simulations using STEMCELL software to analyze electron beam properties. 
Additional setups, such as the Technion UTEM, and synchronization techniques for pump-probe experiments are briefly covered. 

\section{Acknowledgments}
B.M.F. thanks V. Di Giulio, C. Roques-Carmes and K. Beeks for insightful conversations. We thank J.-C. Olaya and A. Hermerschmidt from HOLOEYE for their technical support with the SLM.
This work is part of the SMART-electron project that has received funding from the European Union’s Horizon 2020 Research and Innovation Programme under Grant Agreement No 964591.

\bibliography{pelm_paper} 

\end{document}